\documentclass[%
 aip,
 amsmath,amssymb,
reprint, %
]{revtex4-1}
\usepackage{color}
\usepackage{graphicx}
\usepackage{dcolumn}
\usepackage{bm}
\usepackage{CJK}

\begin{document}


\preprint{AIP/123-QED}

\title{Maximal entropy random walk improves efficiency of trapping in dendrimers}

\author{Xin Peng}

\author{Zhongzhi Zhang}
\email{zhangzz@fudan.edu.cn}
\homepage{http://www.researcherid.com/rid/G-5522-2011}

\affiliation {School of Computer Science, Fudan University,
Shanghai 200433, China}

\affiliation {Shanghai Key Lab of Intelligent Information
Processing, Fudan University, Shanghai 200433, China}

\date{\today}

\begin{abstract}
We use maximal entropy random walk (MERW) to study the trapping problem in dendrimers modeled by Cayley trees with a deep trap fixed at the central node. We derive an explicit expression for the mean first passage time from any node to the trap, as well as an exact formula for the average trapping time (ATT), which is the average of the source-to-trap mean first passage time over all non-trap starting nodes. Based on the obtained closed-form solution for ATT, we further deduce an upper bound for the leading behavior of ATT, which is the fourth power of $\ln N$, where $N$ is the system size. This upper bound is much smaller than the ATT of trapping depicted by unbiased random walk in Cayley trees, the leading scaling of which is a linear function of $N$. These results show that MERW can substantially enhance the efficiency of trapping performed in dendrimers.
\end{abstract}

\pacs{36.20.-r, 05.40.Fb, 05.60.Cd}


\maketitle



\section{Introduction}

Dendrimers are an important class of artificial macromolecules with a treelike structure, which are synthesized by repeating units in a hierarchical self-similar fashion around a central core~\cite{ToNaGo90,ArAs95}. Their unique structural features make them promising candidates for a broad range of potential applications such as light harvesting antennae~\cite{BaCaDeAlSeVe95,KoShShTaXuMoBaKl97} and molecular amplifiers~\cite{AmPeShSh03,SzKeMc03}, the latter of which can be used as efficient platforms for drug delivery~\cite{ShLoSh04}. In view of their practical significance, thus far, dendrimers have received extensive attention within the scientific community~\cite{SuHaFrMu99,SuHaFr00,GaFeRa01,BiKaBl01,MuBiBl06,DoBl09,MuBl11,FuDoBl12,JuWuZh13}.

In the context of light harvesting by dendrimers, it is the large number of absorbing elements at the periphery and an efficient transfer of the absorbed energy to the center that make dendrimers work as antennas~\cite{ShSwShTaXuDcMoKo97}. Generally, light harvesting can be described as a trapping process with a (fluorescent) trap located at the center. A primary quantity related to trapping is average trapping time (ATT), which is the average of mean first-passage time (MFPT)~\cite{Re01,NoRi04,CoBeMo05,CoBeKl07,CoBeMo07,CoBeTeVoKl07} to the target over all starting nodes, where MFPT from a node to the trap is the expected time steps needed for a walker starting off from this node to visit the trap for the first time. ATT is a quantitative indicator measuring the trapping efficiency, which has been much studied for diverse complex systems~\cite{GLKo05,GLLiYoEvKo06,KaRe89,Ag08,LiWuZh10,KaBa02PRE,LiuZh13,ZhQiZhXiGu09,TeBeVo09,AgBu09,AgBuMa10,MeAgBeVo12,HwLeKa12,HwLeKa12E,YaZh13}.\emph{}

Because of the theoretical and practical relevance, trapping in dendrimers has also been devoted to concerted efforts. The problem was first addressed in Refs.~\cite{BaKlKo97,BaKl98,BaKl98JOL}, where the MFPT from a peripheral node to the central node was computed analytically. In Refs.~\cite{BeHoKo03,BeKo06,WuLiZhCh12,LiZh13JCP}, MFPT from any node to the central node, as well as the ATT to the center, were deduced. These works unveiled the effect of structure on the MFPT and ATT to the central node. Note that most previous works on trapping dendrimers based on discrete time random walks focused on the unbiased random walk. However, sometimes it is more suitable to describe some particular problems by a biased random walk than the unbiased one~\cite{ZhShCh13}, since the transition probability depends on not only network topologies but also other properties relevant to the diffusion dynamics~\cite{GoLa08}.

Among numerous biased random walks, maximal entropy random walk (MERW)~\cite{Pa64}, maximizes the entropy of paths, has been studied recently~\cite{BuDuLuWa09}. In MERW, the transition probability incorporates the node centrality measured by the eigenvector associated with the largest eigenvalue of adjacent matrix of the graph where the dynamical process takes place. This local transition rate leads to substantial effects on the diffusion behaviors such as stationary distribution~\cite{BuDuLuWa09} and relaxation time~\cite{OcBu12}. The principle of entropy maximization has found some applications, including optimal sampling algorithm~\cite{He84} and demographic stability of population~\cite{DeGuOc04,DeMa05}. Very recently, as a powerful tool, MERW has been fruitfully applied in the analysis of complex networks~\cite{DeLi11,SiGoLaNiLa11,AnBiSe11,Oc12,OcBu13}. At present, it is still of theoretical and practical interest to explore possible applications of MERW on other areas.

In this paper, we study the trapping problem in Cayley trees as models of dendrimers~\cite{CaCh97,ChCa99}, based on MERW that incorporates the centrality of nodes of the macromolecular graphs. We concentrate on a particular case with a perfect trap placed at the central node, for which we derive an analytical exact expression for ATT. We then provide an upper bound for ATT, whose leading scaling behaves with the system size $N$ as $(\ln N)^4$, a scaling much smaller than that corresponding to unbiased random walk in Cayley trees, for which the ATT scales linearly with $N$. Theses results show that for trapping process in dendrimers with a trap fixed at the center, MERW is considerably more efficient in comparison to unbiased random walk.

\section{Construction and properties of dendrimers}

In this section, we introduce the construction and some relevant features of dendrimers modeled by Cayley trees, which are built in an iterative way. The particular
construction process allows for analytically determining the properties of Cayley trees and deriving exact solutions for diverse dynamical processes on large but finite structures.

\subsection{Construction algorithm}

\begin{figure}
\begin{center}
\includegraphics[width=0.93\linewidth,trim=0 0 0 0]{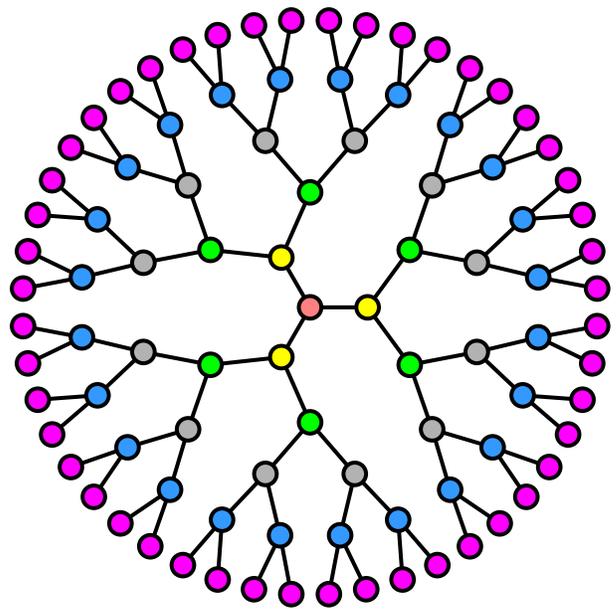}
\caption{(Color online) The Cayley tree $C_{3,5}$. }  \label{Cayley}
\end{center}
\end{figure}

Cayley trees after $n$ iterations (generations), denoted by $C_{m,n}$ ($m\geq 3$, $n \geq 0$) are constructed as follows~\cite{CaCh97,ChCa99}. At the initial generation ($n=0$), $C_{m,0}$ contains only a central node; at $n=1$, $m$ new nodes are generated and linked to the central node to form $C_{m,1}$. These $m$ new single-degree nodes constitute the peripheral nodes of $C_{m,1}$. For any $n>1$, $C_{m,n}$ is obtained from $C_{m,n-1}$: For each peripheral node of $C_{m,n-1}$, $m-1$ new nodes are created and are connected to the peripheral node. Figure~\ref{Cayley} illustrates the structure for a specific dendrimer $C_{3,5}$. Let $N_i$ denote the number of nodes in $C_{m, n}$, which
are created at $i$th generation. Then, we can verify that
\begin{equation} \label{C1}
N_i=
\begin{cases}
1, &{ i=0}\,, \\
m(m-1)^{i-1},&{ i>0}\,. \\
\end{cases}
\end{equation}
Thus, the total number of nodes in $C_{m,n}$ is
\begin{equation} \label{C2}
N=\sum_{i=0}^{n}N_i=\frac{m(m-1)^n-2}{m-2}\,. \\
\end{equation}

\subsection{Largest eigenvalue and its corresponding eigenvector of adjacency matrix}

The special construction of Cayley trees also makes it possible to analytically determine the eigenvalues and their associated eigenvectors~\cite{ZiOb78,HeLiSt03,Sc06,OcBu12}.
Let $\mathbf{A}_n=[A_{ij}]_{N \times N}$
denote the adjacency matrix of network $C_{m,n}$, in which $A_{ij} =A_{ji}=1$
if nodes $i$ and $j$ are linked to each other, $A_{ij} =A_{ji}=0$ otherwise.
Although for a general graph, it is often difficult to determine the
eigenvalues and eigenvectors of its adjacency
matrix, for $C_{m,n}$ the problem can be settled. Note that $C_{m,n}$ has $N$ eigenvalues.  We represent these
$N$ eigenvalues as $\lambda_{1}$,
$\lambda_{2}$, $\dots$, $\lambda_{N-1}$, and $\lambda_{N}$, respectively. Every eigenvalue $\lambda_{i}$ takes the form $\lambda_{i}= 2\sqrt{m-1}\cos \theta_i$  $(i=1,2,\cdots,N)$ where $0<\theta_i<\pi$.

In the sequel, what we are concerned with is the largest eigenvalue and its corresponding eigenvector of $\mathbf{A}_n$, which are denoted by $\lambda$ and $\vec{\psi}$, respectively. The largest eigenvalue $\lambda$ can be expressed as
\begin{equation}
\lambda= 2\sqrt{m-1}\cos \theta\,,
\end{equation}
where
\begin{equation}\label{B0}
\theta \approx \frac{\pi}{n+\frac{2(m-1)}{m-2}}\,.
\end{equation}

With regard to the eigenvector $\vec{\psi}$, since all its entries $\psi_{i}$ for nodes in a given generation $i$ are identical, it can be written as:
\begin{equation}\label{B1}
\vec{\psi}=(\psi_0,\underbrace{\psi_1,\cdots,\psi_1}_{N_1},\cdots,\underbrace{\psi_n,\cdots,\psi_n}_{N_n})
\end{equation}
In order to express the entry $\psi_i$,
we introduce the following quantity:
\begin{equation}\label{B2}
R_i=\frac{\psi_{n-i}}{\psi_n}
\end{equation}
for $i=0,1,2,\cdots,n-1,n$.
It is easy to verify
\begin{equation}\label{B3}
R_i=(m-1)^{i/2}\,\frac{\sin[(i+1)\theta]}{\sin\theta}\,.
\end{equation}
From Eq.~(\ref{B2}), we have
\begin{equation} \label{B4}
\psi_i=R_{n-i}\psi_n\,.
\end{equation}
In the following text, we choose $\psi_n=\sqrt{\sum_{i=0}^{n} [N_i \times(R_{n-i})^2]}$. Thus,
\begin{equation} \label{B5}
\psi_i=\frac{R_{n-i}}{\sqrt{\sum_{i=0}^{n} [N_i \times(R_{n-i})^2]}}\,.
\end{equation}
The selection of $\psi_n$  ensures the proper normalization of every entry of eigenvector $\vec{\psi}$:
\begin{equation} \label{B6}
\sum_{i=0}^{n}[N_i \times(\psi_i)^2]=1\,.
\end{equation}

\section{Maximal entropy random walk in dendrimers with a trap at the center}

After introducing the construction and related properties of Cayley trees $C_{m,n}$, in this section, we study MERW in dendrimers with a perfect trap fixed at the central node. Our goal is to unveil the influence of maximal entropy random walk on the efficiency of trapping performed on this important family of polymer networks.

\subsection{Formulating the problem}

The MERW considered here is a discrete time random walk~\cite{BuDuLuWa09}.
At each discrete time step, the walker jumps from its current position $i$ to any of its neighboring nodes $j$ with probability
\begin{equation} \label{D1}
P_{ij} = \frac{A_{ij}}{\lambda}\frac{\psi_j}{\psi_i} \,.
\end{equation}
Let $\Omega_i$ denote the set of neighbors of a node $i$. Then, for an arbitrary node $i$, $P_{ij}$ fulfills the condition $\sum_{j \in \Omega_i} P_{ij} = 1$, since
\begin{equation}\label{D2}
\sum_{j \in \Omega_i}A_{ij}\psi_j=\lambda\psi_i\,.
\end{equation}
The stationary distribution for MERW on $C_{m,n}$ is~\cite{OcBu12}
\begin{equation}\label{D3}
\pi=(\psi_0^2,\underbrace{\psi_1^2,\cdots,\psi_1^2}_{N_1},\cdots,\underbrace{\psi_n^2,\cdots,\psi_n^2}_{N_n})\,.
\end{equation}

The main subject we focus on here is trapping problem in $C_{m,n}$ described by MERW with a deep trap located at the innermost node. And the main quantity we are interested in is the ATT to the trap. For convenience of description, we label the central node of $C_{m, n}$ by $1$, and consecutively label all other nodes as $2, 3,\cdots, N - 1$, and $N$. We use $F_i(n)$ to represent the trapping time of node $i$, which is defined as the expected time for a walker starting off from node $i$ to reach the trap in $C_{m, n}$ for the first time. Then, the ATT denoted by $\langle F \rangle_n$ is the average of $F_i(n)$ over all non-trap nodes distributed uniformly in $C_{m, n}$, that is,s
\begin{eqnarray} \label{D4}
\langle F \rangle_n = \frac{1}{N-1}\sum_{i=2}^{N}F_i(n)\,.
\end{eqnarray}

Note that MERW in any connected binary network can be represented as generic random walk in a corresponding weighted network~\cite{LaSiDeEvBaLa11}. For $C_{m, n}$, the generalized adjacency matrix (weight matrix) $\mathbf{W}_n=[w_{ij}]_{N \times N}$ corresponding to the weighted networks, denoted by $G_{m, n}$, associated with MERW is defined as follows. The entries $w_{ij}=w_{ji}=\psi_i A_{ij}\psi_j$ if nodes $i$ and $j$ are adjacent in $C_{m, n}$, and   the element $w_{ij}=w_{ji}=0$ if nodes $i$ and $j$ are not directly connected by an edge in $C_{m, n}$. In  weighted networks $G_{m, n}$, the strength~\cite{BaBaVe04} of a node $i$ is defined by $s_i=\sum_{j \in \Omega_i} w_{ij}=\lambda \psi_i^2$, which is $i$th nonzero entry of the diagonal strength matrix $\mathbf{S}_n = {\rm diag}(s_1, s_2, \ldots, s_N)$, and the Laplacian matrix of $G_{m, n}$ is defined to be $\mathbf{L}_n=\mathbf{S}_n-\mathbf{W}_n$.

For generic random walk in $G_{m, n}$, the transition probability for the walker from current state $i$ to one of its neighboring nodes $j$ is $P_{ij}=w_{ij}/s_i=\frac{A_{ij}}{\lambda}\frac{\psi_j}{\psi_i}$, which is the same as that of MERW in $C_{m, n}$. Therefore, the stationary distribution for generic random walk in $G_{m, n}$ is also the same as that corresponding to MERW in $C_{m,n}$.  According to recently obtained result about trapping in weighted networks~\cite{LiZh13}, the explicit expression for ATT $\langle F \rangle_n$ in Eq.~(\ref{D4}) can be expressed as
\begin{equation}\label{D5}
\langle F \rangle_n=\frac{N}{N-1}\sum_{k=2}^N\frac{1}{\eta_k}\left(s\times\mu_{k1}^2-\mu_{k1}\sum_{z=1}^N s_z\mu_{kz}\right)\,.
\end{equation}
In Eq.~(\ref{D5}), $s$ is the sum of strengths over all nodes in $G_{m, n}$, namely $s=\sum_{i=1}^N s_i=\lambda$; $\eta_1, \eta_2, \ldots, \eta_N$ are the $N$ eigenvalues of $\mathbf{L}_n$, rearranged as $0=\eta_1<\eta_2\leq\cdots\leq\eta_N$, and $\mu_1, \mu_2, \ldots, \mu_N$ are the corresponding mutually orthogonal eigenvectors of unit length, where $\mu_i=(\mu_{i1}, \mu_{i2}, \cdots, \mu_{iN})^{\top}$.

It should be mentioned that although the expression for ATT $\langle F \rangle_n$ provided by Eq.~(\ref{D5}) seems compact, it requires computing the eigenvalues and eigenvectors of matrix $\mathbf{L}_n$. Since the network size $N$ grows exponentially with $n$ as shown in Eq.~(\ref{C2}), for moderately large $n$, the computation of spectra demands an impractically large
computational effort. Moreover, by using  Eq.~(\ref{D5}) it is very hard and even impossible to obtain useful information about the dependence of
the leading behavior of $\langle F \rangle_n$ on the network size $N$. It is thus of significant practical importance to seek
other approaches for determining the ATT $\langle F \rangle_n$.
Fortunately, the specific architecture of Cayley trees allows for analytical calculation of $\langle F \rangle_n$ and evaluation of its dominant scaling.

\subsection{Exact solution to average trapping time}

According to the structure of dendrimers, all the nodes in $C_{m, n}$ can be classified into $n+1$ levels. The central node is at level $0$, and nodes born at the first generation $i$ are at level $1$, and so forth. By symmetry, 
the trapping time for nodes at the same level is identical. In the case without confusion, we use $F_i(n)$ to denote the trapping time for a node at $i$th level. Then, the following relations hold:
\begin{small}
\begin{equation}\label{C10}
F_i(n)=
\begin{cases}
  0,&{i=0},\\
P_i^{\rm{up}}[1+F_{i-1}(n)]+P_i^{\rm{down}}[1+F_{i+1}(n)],&{0<i<n},\\
1+F_{i-1}(n),&{i=n}.
\end{cases}
\end{equation}
\end{small}
In Eq.~(\ref{C10}), $P_i^{\rm{up}}$ represents the transition probability for a walker hopping from a node at level $i$ to its (unique) father node at level $i-1$, while $P_i^{\rm{down}}$ denotes the transition probability from any node at level $i$ to its $m-1$ child nodes at level $i+1$. For unbiased random walk in $C_{m,n}$~\cite{WuLiZhCh12,LiZh13JCP}, $P_i^{\rm{up}}=\frac{1}{m}$ and  $P_i^{\rm{down}}=\frac{m-1}{m}$, obeying relation $P_i^{\rm{up}}+P_i^{\rm{down}}=1$. However, as will be shown below, for MERW in $C_{m,n}$, $P_i^{\rm{up}}$ and $P_i^{\rm{down}}$ are different from those corresponding to unbiased random walk, although $P_i^{\rm{up}}+P_i^{\rm{down}}=1$ also holds. And the disparity for transition probability between these two kinds of random walks leads to quite different scalings for ATT.


By definition of transition probability for MERW in $C_{m,n}$, see Eq.~(\ref{D1}), $P_i^{\rm{up}}$ and $P_i^{\rm{down}}$ for MERW in $C_{m,n}$ are
\begin{equation} \label{C11}
P_i^{\rm{up}} =\frac{1}{\lambda}\frac{\psi_{i-1}}{\psi_i}
\end{equation}
and
\begin{equation} \label{C12}
P_i^{\rm{down}} =(m-1)\frac{1}{\lambda}\frac{\psi_{i+1}}{\psi_i}\,,
\end{equation}
respectively. Considering above-obtained results, $P_i^{\rm{up}}$ and $P_i^{\rm{down}}$ can be further recast as
\begin{eqnarray} \label{C16}
P_i^{\rm{up}}&=&\frac{1}{\lambda}\frac{R_{n-i+1}\psi_n}{R_{n-i}\psi_n} \nonumber \\
&=&\frac{1}{2\sqrt{m-1}\cos\theta}\frac{(m-1)^{(n-i+1)/2}\frac{\sin[(n-i+1+1)\theta]}
{\sin\theta}}{(m-1)^{(n-i)/2}\frac{\sin[(n-i+1)\theta]}{\sin\theta}} \nonumber \\
&=&\frac{1}{2\cos\theta}\frac{\sin[(n-i+2)\theta]}{\sin[(n-i+1)\theta]} \nonumber\\
&=&\frac{\sin[(n-i+2)\theta]}{\sin[(n-i+2)\theta]+\sin[(n-i)\theta]}
\end{eqnarray}
and
\begin{eqnarray} \label{C18}
P_i^{\rm{down}}&=&(m-1)\frac{1}{\lambda}\frac{R_{n-i-1}\psi_n}{R_{n-i}\psi_n} \nonumber \\&=&(m-1)\frac{1}{2\sqrt{m-1}\cos\theta}\frac{(m-1)^{(n-i-1)/2}\frac{\sin[(n-i)\theta]}
{\sin\theta}}{(m-1)^{(n-i)/2}\frac{\sin[(n-i+1)\theta]}{\sin\theta}} \nonumber\\
&=&\frac{1}{2\cos\theta}\frac{\sin[(n-i)\theta]}{\sin[(n-i+1)\theta]} \nonumber \\
&=&\frac{\sin[(n-i)\theta]}{\sin[(n-i+2)\theta]+\sin[(n-i)\theta]}\,,
\end{eqnarray}
both of which evidently fulfil the relation $P_i^{\rm{up}}+P_i^{\rm{down}}=1$, implying that our computation for $P_i^{\rm{up}}$ and $P_i^{\rm{down}}$ is correct.

Using Eqs.~(\ref{C16}) and~(\ref{C18}), Eq.~(\ref{C10}) becomes
\begin{equation} \label{C19}
F_i(n)=
\begin{cases}
  0,&{i=0},\\
\frac{\sin[(n-i+2)\theta][1+F_{i-1}(n)]}{\sin[(n-i+2)\theta]+\sin[(n-i)\theta]}  \\
+\frac{\sin[(n-i)\theta][1+F_{i+1}(n)]}{\sin[(n-i+2)\theta]+\sin[(n-i)\theta]},
&{0<i<n}, \\
1+F_{i-1}(n),&{i=n}.\\
\end{cases}
\end{equation}
Therefore, for $0<i<n$, we have
\begin{eqnarray} \label{C20}
&&{\{\sin[(n-i+2)\theta]+\sin[(n-i)\theta]\}}F_i(n)\nonumber \\
&=&\sin[(n-i+2)\theta][1+F_{i-1}(n)]\nonumber \\
&&+\sin[(n-i)\theta][1+F_{i+1}(n)]\,,
\end{eqnarray}
which can be rewritten as
\begin{eqnarray} \label{C21}
&&\sin[(n-i+2)\theta][F_i(n)-F_{i-1}(n)]\nonumber \\
&=&\sin[(n-i+2)\theta]+\sin[(n-i)\theta] \nonumber \\
&&+\sin[(n-i)\theta][F_{i+1}(n)-F_i(n)]\,.
\end{eqnarray}

In order to determine the quantity $\langle{F}\rangle_n$, we define
\begin{equation} \label{C22}
B_i(n) = F_{n-i}(n)-F_{n-i-1}(n)\,.
\end{equation}
Then,
\begin{equation} \label{C23}
B_i(n)=1+\frac{\sin{i\theta}}{\sin[(i+2)\theta]}+\frac{\sin{i\theta}}{\sin[(i+2)\theta]}B_{i-1}(n) \end{equation}
holds for all $0<i<n$. Note that
\begin{eqnarray} \label{C24}
B_0(n)&=&F_n(n)-F_{n-1}(n) \nonumber \\
&=& [1+F_{n-1}(n)]-F_{n-1}(n) \nonumber \\
&=&1\,,
\end{eqnarray}
using which Eq.~(\ref{C23}) can be solved to yield
\begin{equation} \label{C25}
B_i(n)=2\sum_{j=0}^{i-1}\prod_{k=0}^j\frac{\sin[(i-k)\theta]}{\sin[(i+2-k)\theta]}+1.
\end{equation}
Making use of Eqs.~(\ref{C22}) and~(\ref{C25}), $F_i(n)$ can be evaluated as
\begin{eqnarray} \label{C26}
F_i(n)&=&B_{n-i}(n)+F_{i-1}(n) \nonumber \\
&=&B_{n-i}(n)+B_{n-i+1}(n)+F_{i-2}(n) \nonumber \\
&=&\cdots \cdots=\sum_{j=0}^{i-1}B_{n-i+j}(n)+F_0(n)=\sum_{j=1}^iB_{n-j}(n) \nonumber \\
&=&\sum_{j=1}^i\Bigg[2\sum_{k=0}^{n-j-1}\prod_{l=0}^k\frac{\sin[(n-j-l)\theta]}{\sin[(n-j+2-l)\theta]}+1\Bigg]\,,
\end{eqnarray}
where $F_0(n)=0$ was used. Equation~(\ref{C26}) provides a closed form expression for trapping time for an arbitrary node.

\begin{figure}
\begin{center}
\includegraphics[width=1.1\linewidth,trim=30 50 0 55]{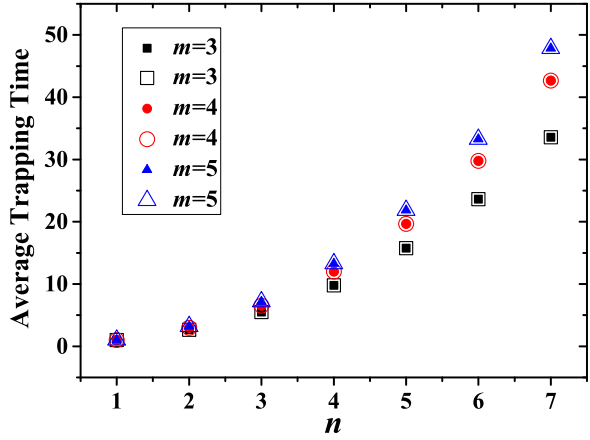}
\end{center}
\caption[kurzform]{\label{trapping} (Color online) The ATT $\langle F\rangle_n$ to the central node corresponding to MERW in $C_{m,n}$ for various $m$ and $n$. The hollow symbols represent the results generated by Eq.~(\ref{D5}), while the solid symbols stand for the analytical results given by Eq.~(\ref{C27}).}
\end{figure}

Inserting Eq.~(\ref{C26}) into Eq.~(\ref{D4}), we can obtain the closed-form formula of ATT $\langle{F}\rangle_n$ for MERW in $C_{m,n}$ with a single trap positioned at the central node, which reads
\begin{widetext}
\begin{eqnarray} \label{C27}
\langle{F}\rangle_n&=&\frac{1}{N-1}\displaystyle{\sum_{j=2}^{N}}F_j(n)=
\frac{\sum_{i=1}^n[N_i \times F_i(n)]}{N-1} \nonumber \\
&=&\frac{\displaystyle{\sum_{i=1}^n}\left\{m(m-1)^{i-1}\displaystyle{\sum_{j=1}^i}\left[2\displaystyle{\sum_{k=0}^{n-j-1}}\displaystyle{\prod_{l=0}^k}\frac{\sin[(n-j-l)\theta]}{\sin[(n-j+2-l)\theta]}
+1\right]\right\}}{\frac{m[(m-1)^n-1]}{m-2}}\nonumber \\
&=&\frac{m-2}{(m-1)^n-1}\sum_{i=1}^n \left\{(m-1)^{i-1}\sum_{j=1}^i
\left[2\sum_{k=0}^{n-j-1}\prod_{l=0}^k\frac{\sin[(n-j-l)\theta]}{\sin[(n-j+2-l)\theta]}+1
 \right]\right\}\,.
\end{eqnarray}
\end{widetext}
In Fig.~\ref{trapping}, we present results about the ATT $\langle{F}\rangle_n$ to the central node for MERW in $C_{m,n}$. The results are obtained by using Eqs.~(\ref{D5}) and~(\ref{C27}), respectively. Figure~\ref{trapping} shows that the results generated by Eq.~(\ref{D5}) and Eq.~(\ref{C27}) agree with each other, confirming our theoretical results for $\langle{F}\rangle_n$ provided by Eq.~(\ref{C27}).

\subsection{An upper bound for leading scaling of average trapping time}

Although the expression for the ATT $\langle{F}\rangle_n$ given in Eq.~(\ref{C27}) is exact, it is rather lengthy and awkward, from which we cannot see obvious dependence of $\langle{F}\rangle_n$ on the network size $N$. However, we will show below that, when the networks are large enough, from Eq.~(\ref{C27}) one can derive an upper bound for the leading scaling of $\langle{F}\rangle_n$ in terms of the network size $N$.

First, let us examine the product term
\begin{equation} \label{C28}
\prod_{l=0}^k\frac{\sin[(n-j-l)\theta]}{\sin[(n-j+2-l)\theta]}
\end{equation}
in Eq.~(\ref{C27}). Seemingly, there are $k+1$ multipliers in both the denominator and numerator in Eq.~(\ref{C28}). In fact, for any $k$ we can decrease the number of multipliers from $k+1$ to two by simplifying Eq.~(\ref{C28}) as follows:
\begin{small}
\begin{eqnarray} \label{C29}
&&\prod_{l=0}^k\frac{\sin[(n-j-l)\theta]}{\sin[(n-j+2-l)\theta]} \nonumber \\
&=&\frac{\sin[(n-j)\theta]\sin[(n-j-1)\theta]\cdots\sin[(n-j-k)\theta]}{\sin[(n-j+2)\theta]\sin[(n-j+1)\theta]\cdots\sin[(n-j+2-k)\theta]} \nonumber \\
&=&\frac{\sin[(n-j+1-k)\theta]\sin[(n-j-k)\theta]}{\sin[(n-j+2)\theta]\sin[(n-j+1)\theta]}\,.
\end{eqnarray}
\end{small}

Since $\theta<\frac{\pi}{n+2}$, for any $1\leq j \leq i$ ($1\leq i \leq n$) and $0\leq l \leq k$ ($0\leq k \leq n-j-1$), every term in both the denominator and numerator of Eq.~(\ref{C29}) has the form $\sin x\theta$, where  $x$ is an integer between $1$ and $n+1$. Thus, $x\theta \in (0, \pi)$ and $0<\sin x\theta\le1$. According to Eq.~(\ref{B0}), we have
\begin{equation} \label{C30}
x\theta \approx \frac{x\pi}{n+\frac{2(m-1)}{m-2}}>\frac{1}{cn}\,,
\end{equation}
where $c$ is a certain positive constant. Note that if constant $c$ is chosen properly, the following relation also holds:
\begin{eqnarray} \label{C31}
\pi-x\theta&\geq&\pi-\frac{(n+1)\pi}{n+\frac{2(m-1)}{m-2}} \nonumber \\
&=&\pi\frac{[(m-2)n+2m-2]-[(n+1)(m-2)]}{(m-2)n+2m-2} \nonumber \\
&>&\frac{1}{cn}\,.
\end{eqnarray}
On the other hand, when $n$ is large enough, $\sin \frac{1}{cn} \approx \frac{1}{cn}$. Then, each term $\sin[(n-j-l)\theta]$ in the numerator of Eq.~(\ref{C28}) satisfies $1\ge\sin[(n-j-l)\theta]>\sin \frac{1}{cn} = \frac{1}{cn}$. Thus,
\begin{equation}\label{C32}
\prod_{l=0}^k\frac{\sin[(n-j-l)\theta]}{\sin[(n-j+2-l)\theta]}\le \frac{1}{\frac{1}{cn}} \frac{1}{\frac{1}{cn}} \approx d\,n^2,
\end{equation}
where $d=c^2$ is another positive constant.

Substituting the result in Eq.~(\ref{C32}) into  Eq.~(\ref{C27}), we obtain
\begin{eqnarray} \label{C33}
\langle F\rangle_n&\le&\frac{m-2}{(m-1)^n-1}\sum_{i=1}^n\Bigg\{(m-1)^{i-1} \times\nonumber \\
&& \sum_{j=1}^i\left[2\sum_{k=0}^{n-j-1}dn^2
+1 \right]\Bigg\}
\nonumber \\
&\le&\frac{m-2}{(m-1)^n-1}\sum_{i=1}^n(m-1)^{i-1}dn^4 \nonumber \\
&\le&\frac{m-2}{(m-1)^n-1}\frac{(m-1)^n-1}{m-2}dn^4 \nonumber \\
&\le&d\,n^4\,.
\end{eqnarray}
Note that, Eq.~(\ref{C2}) enables us to represent $n$ in terms of the system size $N$ as
\begin{equation}\label{C34}
n=\frac{\ln[(m-2)N+2]-\ln m}{\ln(m-1)}\,.
\end{equation}
Hence, the upper bound for $\langle F\rangle_n$ in Eq.~(\ref{C33}) can be expressed
in the following form:
\begin{eqnarray} \label{C35}
\langle F\rangle_n &\le& d \left[\frac{\ln[(m-2)N+2]-\ln m}{\ln(m-1)}\right]^4 \nonumber \\
&\le& f(\ln N)^4
\end{eqnarray}
where $f$ is a certain positive constant.

Equation~(\ref{C35}) provides an upper bound of dominant behavior of $\langle F\rangle_n$ for MERW in massive networks $C_{m,n}$ with the immobile trap located at the central node, which shows that the efficiency of trapping in dendrimers described by MERW is very high. It is in sharp contrast with the ATT for unbiased random walk in the same networks, where the leading term of ATT increases lineally with the system size~\cite{WuLiZhCh12,LiZh13JCP}. Thus, MERW is instrumental in improving the efficiency  of trapping process taking place in dendrimers.

\section{Conclusions}

Based on maximal entropy random walk (MERW) we have performed an analytical research on trapping in dendrimers with a single trap placed at the central node. We have derived an explicit expression for ATT as an indicator of the trapping efficiency, grounded on which we have deduced an upper bound for the leading scaling of ATT. The obtained upper bound scales with the network size $N$ as $(\ln N)^4$, and is considerably lower than the ATT for trapping in dendrimers based on unbiased random walk, which grows lineally with the network size $N$. This means that in dendrimers MERW can dramatically lessen the ATT to the central node, making the diffusion process highly efficient. This theoretical work might prove useful to address light harvesting by dendrimers, as well as various other dynamical processes occurring in nanoscale systems.

\begin{acknowledgments}
The authors thank Bin Wu and Yuan Lin for their assistance in preparing this manuscript. This work was supported by the National Natural Science Foundation of China under Grant No. 11275049.
\end{acknowledgments}



\nocite{*}

\end{document}